\documentclass{article}  
\usepackage{lajolla2006}
\usepackage{graphicx}
\usepackage{xspace}
\usepackage[tight]{subfigure}

\newcommand{\pT}{\ensuremath{p_T}\xspace}
\newcommand{\vTwo}{\ensuremath{v_2}\xspace}
\newcommand{\vTwoRaw}{\ensuremath{v_2^{raw}}\xspace}
\newcommand{\vTwoCorr}{\ensuremath{v_2^{corr}}\xspace}
\newcommand{\piz}{\ensuremath{\pi^0}\xspace}
\newcommand{\RAA}{\ensuremath{R_{AA}}\xspace}
\newcommand{\photon}{\ensuremath{\gamma}\xspace}
\newcommand{\meanNpart}{\ensuremath{\langle N_{part}\rangle}\xspace}
\newcommand{\dphi}{\ensuremath{\Delta\phi}\xspace}
\newcommand{\leff}{\ensuremath{\rho~L~dL}\xspace}
\makeatletter
\newcommand*{\SubLabels}[1]{%
  \label{fig:#1}%
  \begingroup
    \protected@edef\@currentlabel{%
      \csname thesub\@captype\endcsname
    }%
    \label{subfig:#1}%
  \endgroup
}
\makeatother

\frompage{000} \topage{000}                                              %|
\def\rightmark{High-$p_{T}$ Production w.r.t. Reaction Plane}
\def\leftmark{D. Winter}
 
\title{High-$p_{T}$ Particle Production with Respect to the Reaction Plane} 
\authors{
{David L. Winter$^1$, {\it for the PHENIX Collaboration}%
}\\[2.812mm]
{\normalsize
\hspace*{-8pt}$^1$ Columbia University, \\ 
New York, NY, USA\\[0.2ex] 
}}
 
\abstract{The PHENIX Run4 data-set provides a powerful opportunity for
exploring the angular anisotropy of identified particle yields at high
$p_{T}$. Complementing traditional $v_2$ measurements, we present
$\pi^0$ yields as a function of emission angle with respect to the
reaction plane in Au+Au collisions at $\sqrt{s_{NN}} = 200$~GeV/c.  The
centrality dependence of the angular anisotropy allows us to probe the
density and path-length dependence of the energy loss of
hard-scattered partons.}

 
\begin{document}
 
\maketitle
\setcounter{page}{1}

\section{Introduction}\label{intro}

Two of the greatest mysteries that have arisen from the RHIC physics program
are the source of the apparent flatness of the high \pT ($>5$~GeV/c)
suppression of \RAA~\cite{ref:tadaaki_pizRaa} and the source of non-zero \vTwo
at high \pT~\cite{ref:highptv2}.  The existence of intermediate to high \pT
\vTwo was suggested early in the RHIC program~\cite{ref:gvw}, and has been the
subject of many theoretical treatments (see~\cite{ref:shuryak,ref:dfj} for
some additional examples).  Traditional flow and parton energy loss pictures
have failed to describe the magnitude of this anisotropy.  Measurement of the
azimuthal asymmetry $v_2$ at high $p_T$ will shed light on the contributions
from flow, recombination, and energy loss, as well as the transition from soft
to hard production mechanisms.
 
\section{Measuring \vTwo and \piz yields in PHENIX}

The orientation of the reaction plane is measured event-by-event using the set
of two PHENIX Beam-Beam Counters (BBCs), which reside at the region
$3<|\eta|<4$.  Each detector is an array of 64 hexagonal, close-packed quartz
Cherenkov counters, located 150 cm from the interaction point.  The charge
measured by each counter is proportional (on average) to the multiplicity of
particles hitting it.  The reaction plane angle $\Psi_{RP}$ is determined from
the value of $\langle cos2\phi \rangle$.  Because the two BBCs provide
independent measurements of $\Psi_{RP}$, we can estimate the resolution of the
combined measurement via standard techniques~\cite{ref:resolution}.

For measuring photons and \piz{}s, we use the Electomagnetic Calorimeter
(EMCal)~\cite{ref:emcal}. Candidate clusters are required to pass \photon
identification cuts, and $m_{inv}$ distributions are formed from pairs of
these clusters.  The resulting yields are binned in angle with respect to the
reaction plane ($\Delta\phi = \phi - \Psi_{RP}$).  A similarly binned mixed
event background is then subtracted. The counts in the remaining peak centered
on the $\pi$ mass are integrated in a $\pm 2\sigma$ window (where $\sigma$ is
the width of a Gaussian fit to the peak).  Six bins in $\Delta\phi$ are used in
the interval $[0-\pi/2]$.

To measure \vTwo, we fit the $\Delta\phi$ distribution as $1 + 2\vTwoRaw
\cos(2\Delta\phi)$.  The resulting \vTwo parameter needs to be corrected for
the reaction plane measurement resolution, hence the designation \vTwoRaw.
The resolution $\sigma_{RP}$ is determined for each centrality bin, and leads
to the corrected value $\vTwoCorr = \vTwoRaw/\sigma_{RP}$.  The yields as a
function of $\Delta\phi$ can then be corrected with a factor
\begin{equation}
f(\Delta\phi) = \frac{1+2\vTwoCorr\cos2\Delta\phi}{1+2\vTwoRaw\cos2\Delta\phi}.
\end{equation}

\section{Results and Discussion}

\begin{figure}[t]
\begin{center}
  \subfigure[\SubLabels{raa_npart}\piz \RAA
    vs. $N_{part}$]{\includegraphics[width=0.65\textwidth]{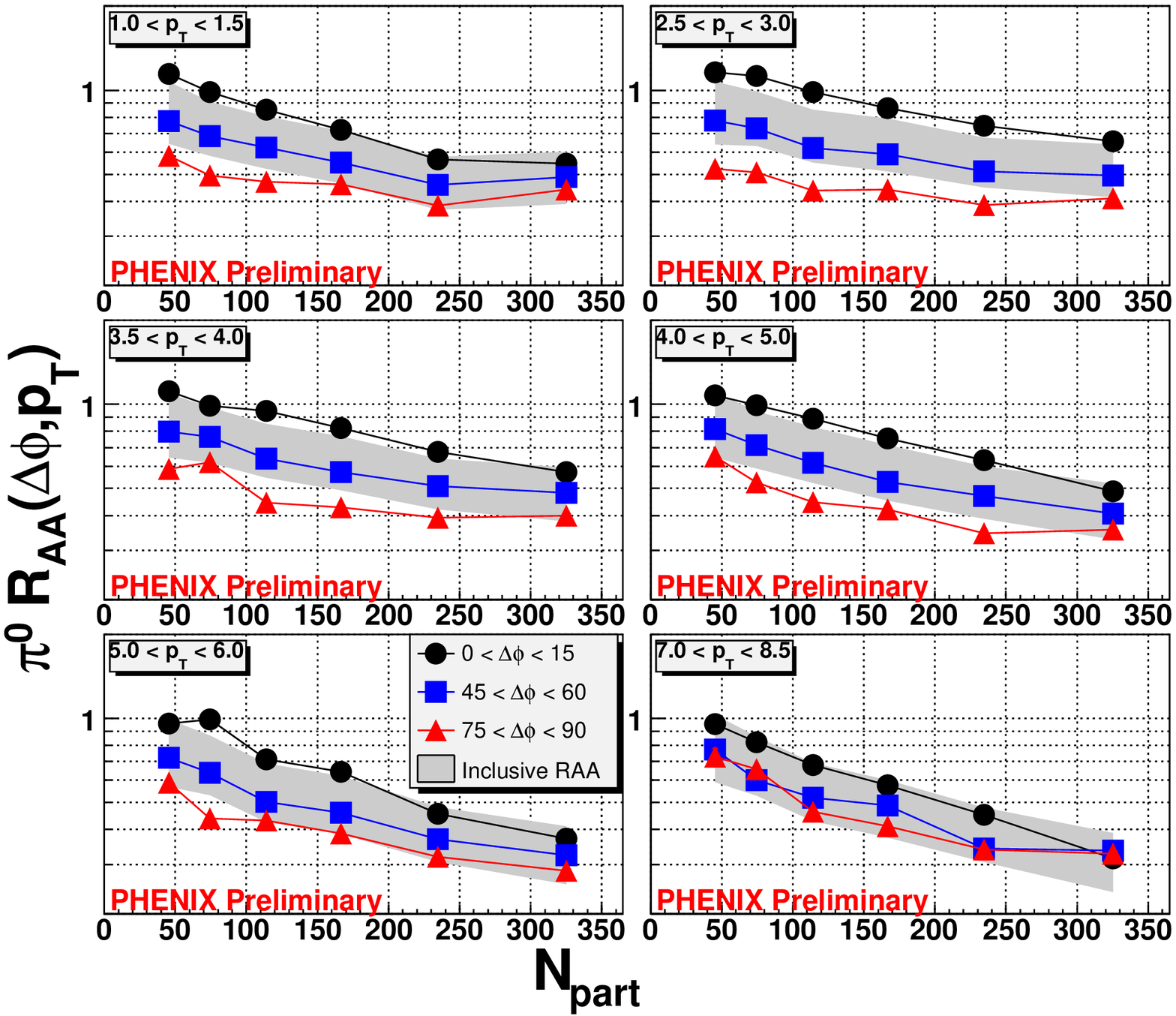}
  }
  \subfigure[\SubLabels{piz_v2}\piz \vTwo
    vs. \pT]{\includegraphics[width=0.65\textwidth]{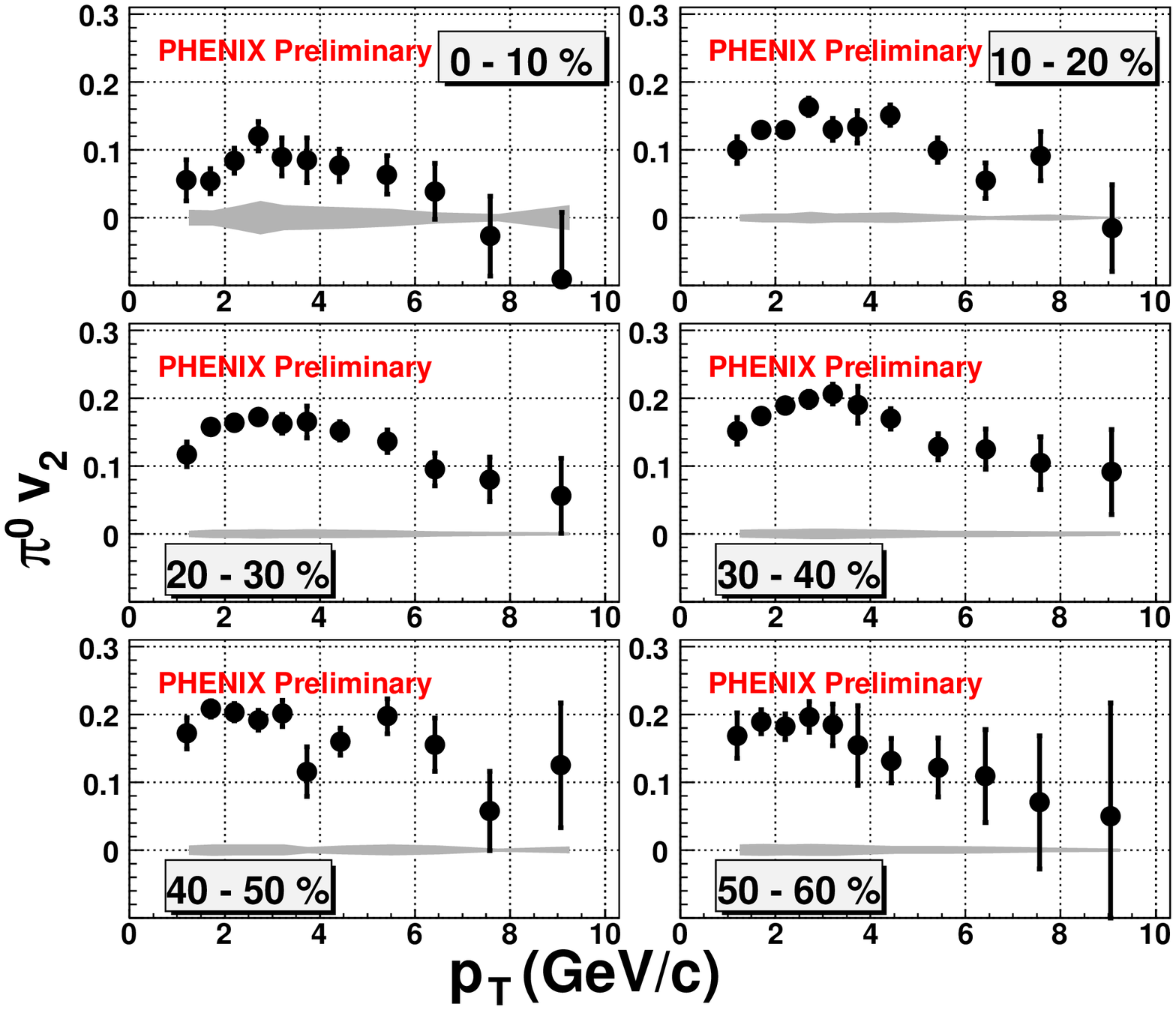}
  }
\end{center}
\vspace{-20pt}
\caption{\piz  \RAA and \vTwo: \ref{subfig:raa_npart} shows $\RAA(\pT)$
  as a function of \meanNpart for each \dphi bin; the panels are for
  different \pT bins.  \ref{subfig:piz_v2} Shows the \piz $\vTwo(\pT)$, with
  each panel corresponding to a centrality bin.  The grey bands indicate the
  systematic error due to the reaction plane resolution correction.}
\end{figure}

To obtain $\RAA(\Delta\phi)$, we exploit the fact that the ratio of the yield
at a given $\Delta\phi$ to the inclusive yield is equivalent to the ratio of
the angle-dependent \RAA to the inclusive \RAA.  Thus multiplying these
relative yields by an inclusive measured $\RAA$, we have:
\begin{equation}
\RAA(\Delta\phi) = {\tt Yield}(\Delta\phi) / {\tt Yield} \times \RAA
\end{equation}

The $\RAA(\Delta\phi,\pT)$ as a function of \meanNpart is shown in
Figure~\ref{fig:raa_npart}.  It is clear that there is non-trivial angular
substructure of the \RAA, and that it varies with centrality.  This feature
is emphasized by plotting the data on a semi-log scale, showing that the \RAA
behaves differently in different \dphi bins.

The resulting \piz \vTwo is shown in Figure~\ref{fig:piz_v2}.  For the first
time we observe \vTwo up to 10 GeV/c.  While the value of the \vTwo decreases
beyond intermediate \pT, it nonetheless shows a substantial and
perhaps constant value out to the highest measured transverse momenta.

\begin{figure}[t]
\begin{center}
\includegraphics*[width=\textwidth]{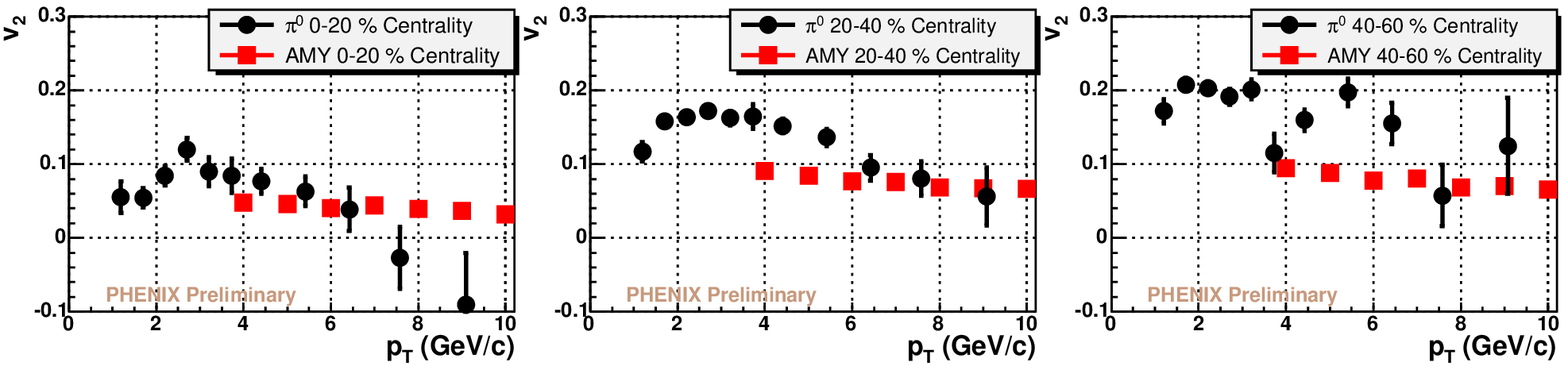}
\includegraphics*[height=4cm]{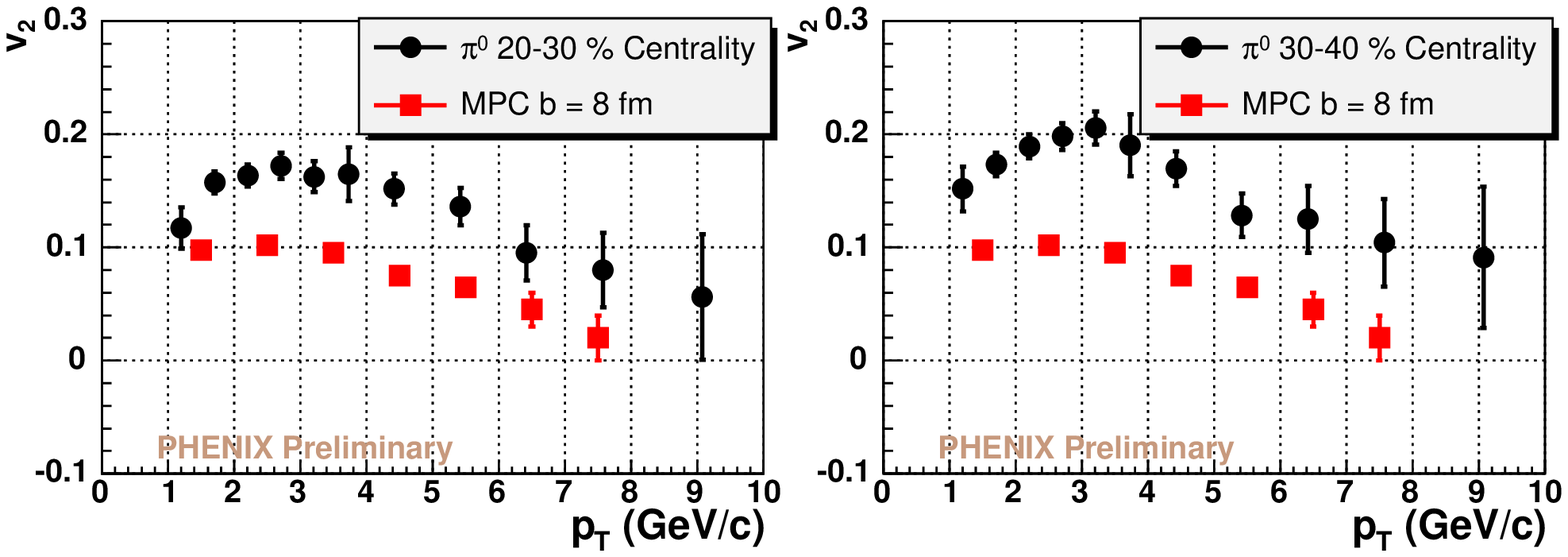}
\end{center}
\caption{Comparison of \piz \vTwo with models. The top three panels show the
  AMY calculation with data for three centralities~\cite{ref:turbide}.  The
  bottom two panels compare two centralities with the $b=8$~fm calculation of
  the MPC model~\cite{ref:mpc}.}
\label{fig:modelCompare}
\end{figure}

To gain insight into the \vTwo mechanisms at work at high \pT, we turn to
models.  We compare the \piz \vTwo to two models, a calculation done by
Turbide et al.~\cite{ref:turbide} (using an Arnold-Moore-Yaffe (AMY)
formalism~\cite{ref:amy}) and the Molnar Parton Cascade (MPC)
model~\cite{ref:mpc}.  Figure~\ref{fig:modelCompare} shows calculations from
these models, plotted alongside data for similar centralities.  The AMY
calculation contains energy loss mechanisms only, and we see that the data
appear to decrease to a value at high \pT that is consistent with this model;
the level of agreement is most striking in the 20-40\% bin.

The MPC model has a number of mechanisms, including corona effects, energy
loss, and the ability to boost lower \pT partons to higher \pT (a unique
feature).  The calculation shown in Figure~\ref{fig:modelCompare} does a
better job of reproducing the overall shape of the \vTwo, though it is
systematically low.  It is important to note that this calculation is done for
one set of parameters, so it should be very interesting to see if the MPC can
better reproduce the data for a different set of parameters (the opacity of
the medium, for example).

The prevailing thought is that the high \pT behavior of the \vTwo is due to
energy loss mechanisms.  If this is true, the \RAA should be sensitive to the
geometry of the collision.  To test this behavior, we seek to reparamaterize
the two handles we have on geometry (centrality, or collision overlap, and
angle of emission) into a single parameter, a quantity which we will refer to
as ``\leff''.  Details of the calculation are described in~\cite{ref:leff}.
This effective path length is calculated from the parton-density weighted
average of the length from hard-scattering origin to edge of an ellipse and
includes the Bjorken 1-D longitudinal expansion.  We perform a Glauber Monte
Carlo sampling of starting points to account for fluctuations in the location
of the hard-scattering origin of the particles' paths within the region of
overlap between the colliding nuclei.  The crucial feature of \leff is that it
is proportional to the energy loss sustained by the parton as it traverses the
medium.

\begin{figure}[t]
\begin{center}
\includegraphics*[width=\textwidth]{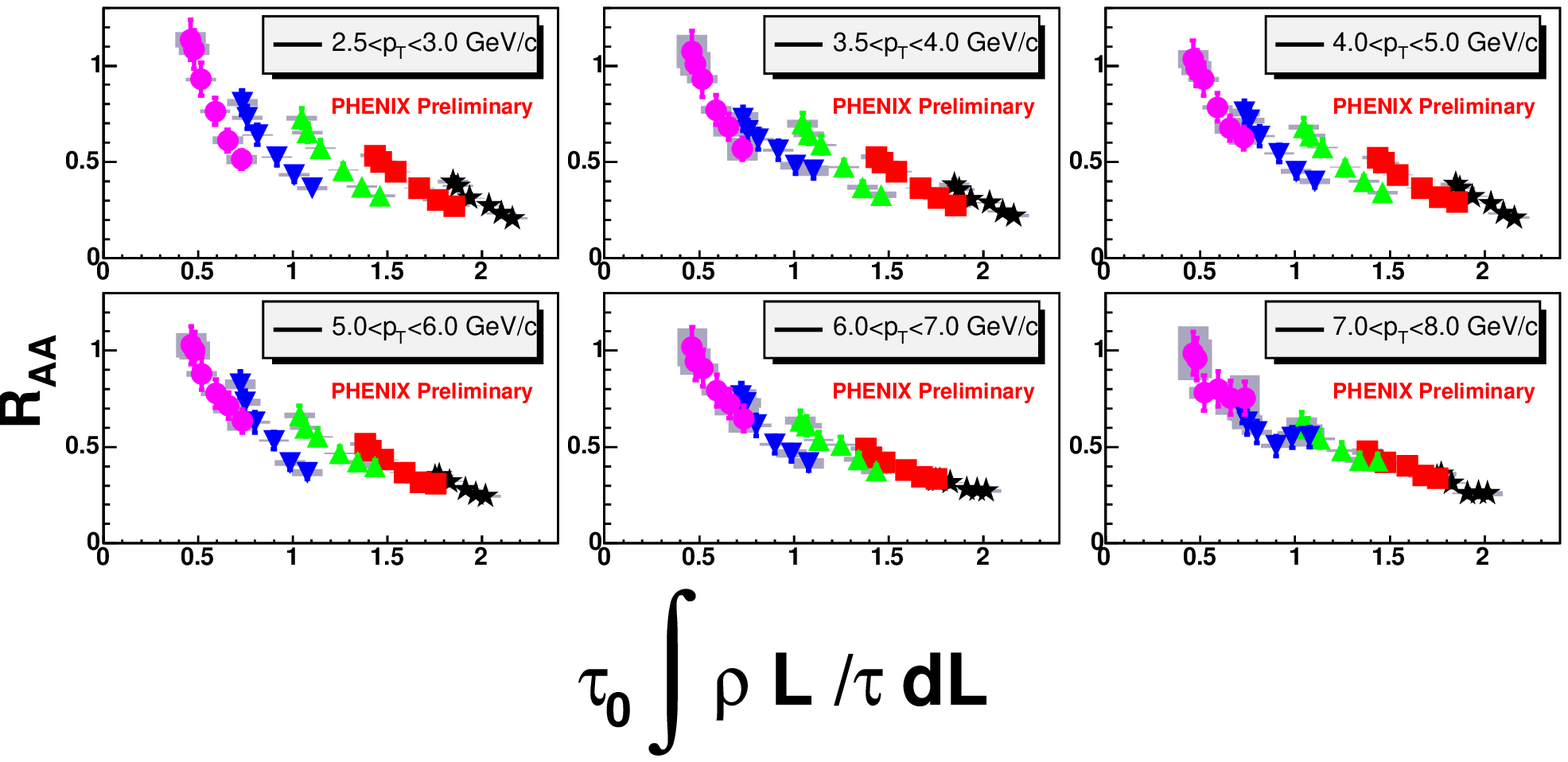}
\end{center}
\vspace{-20pt}
\caption{$R_{AA}(\dphi,\pT)$ vs. $\leff$.  The panels correspond to different
  \pT ranges.  The solid circles are the most peripheral events, while the
  solid stars are the most central events.}
\label{fig:leff}
\end{figure}

The result of plotting \RAA for all centralities and angles
vs. \leff is shown in Figure~\ref{fig:leff}.  If the observed
\RAA arose from only geometric effects, we would expect the data to
exhibit a universal dependence on \leff.  For low \pT, this is clearly
not the case; something more than just energy loss is taking place
there.  However, when the \pT reaches 7 GeV/c and above, the \RAA data
do indeed appear to have a dependence on a single \leff curve.  This apparent
scaling strongly suggests that the dominant effect on \RAA at high-\pT is
radiative energy loss.

\section{Conclusions}\label{concl}
We have presented the first measurement of high \pT \vTwo for \piz{}s.  It is
now clear that the \vTwo at high \pT does decrease but to a non-zero value.
Comparison of \vTwo with models suggest that the dominant mechanism at work at
high \pT is energy loss.  In addition, we have presented the first measurement
of \piz \RAA as a function of angle with respect to the reaction plane.  The
$\RAA(\dphi,\pT)$ exhibits interesting angular substructure.  Furthermore,
when the \RAA data are plotted as a function of an effective path length
through the medium, the scaling that arises at high \pT also argues for energy
loss as the dominant mechanism at work.

\vfill\eject
\end{document}